\documentclass[prl,superscriptaddress,showpacs,twocolumn]{revtex4}

\usepackage{epsfig}
\usepackage{graphics}
\usepackage{amsmath,amsfonts,amssymb}
\def\be{\begin{equation}}
\def\ee{\end{equation}}
\def\bea{\begin{eqnarray}}
\def\eea{\end{eqnarray}}
\def\bml{\begin{mathletters}}
\def\eml{\end{mathletters}}
\def\bc{\begin{center}}
\def\ec{\end{center}}
\def\bn{\bar{n}}

\begin{document}

\title{Percolation, Morphogenesis, and Burgers Dynamics
in Blood Vessels Formation}

\author{A. Gamba}
\affiliation{Dipartimento di Matematica, Politecnico di Torino,
10129 Torino, Italia}
\author{D. Ambrosi}
\affiliation{Dipartimento di Matematica, Politecnico di Torino,
10129 Torino, Italia}
\author{A. Coniglio} 
\affiliation{Dipartimento di Scienze Fisiche,
Universit\`a di Napoli ``Federico II'', and INFM, Unit\`a di Napoli,
80126 Napoli, Italia}
\author{A. de Candia} 
\affiliation{Dipartimento di Scienze Fisiche,
Universit\`a di Napoli ``Federico II'', and INFM, Unit\`a di Napoli,
80126 Napoli, Italia}
\author{S. Di Talia}
\affiliation{Dipartimento di Scienze Fisiche,
Universit\`a di Napoli ``Federico II'', and INFM, Unit\`a di Napoli,
80126 Napoli, Italia}
\author{E. Giraudo} 
\affiliation{Istituto per la Ricerca e la Cura del Cancro,
10060 Candiolo (TO), Italia}
\author{G. Serini}
\affiliation{Istituto per la Ricerca e la Cura del Cancro,
10060 Candiolo (TO), Italia}
\author{L. Preziosi} 
\affiliation{Dipartimento di Matematica, Politecnico di Torino,
10129 Torino, Italia}
\author{F. Bussolino}
\affiliation{Istituto per la Ricerca e la Cura del Cancro,
10060 Candiolo (TO), Italia}

\begin{abstract}
Experiments of {\em in vitro} formation of blood vessels show
that cells randomly spread on a gel matrix autonomously organize
to form a connected vascular network.
We propose 
a simple model which reproduces many features of the biological
system. We
show that both the model and the real system exhibit a fractal 
behavior at small scales, due to the process of migration and dynamical
aggregation, followed at large scale by a random percolation behavior
due to the coalescence of aggregates.
The results are in good agreement with the analysis performed on the
experimental data.
\end{abstract}


\maketitle

\begin{figure}
\bc
\begin{tabular}{cccc}
a&\mbox{\epsfysize=2.5cm\epsfbox{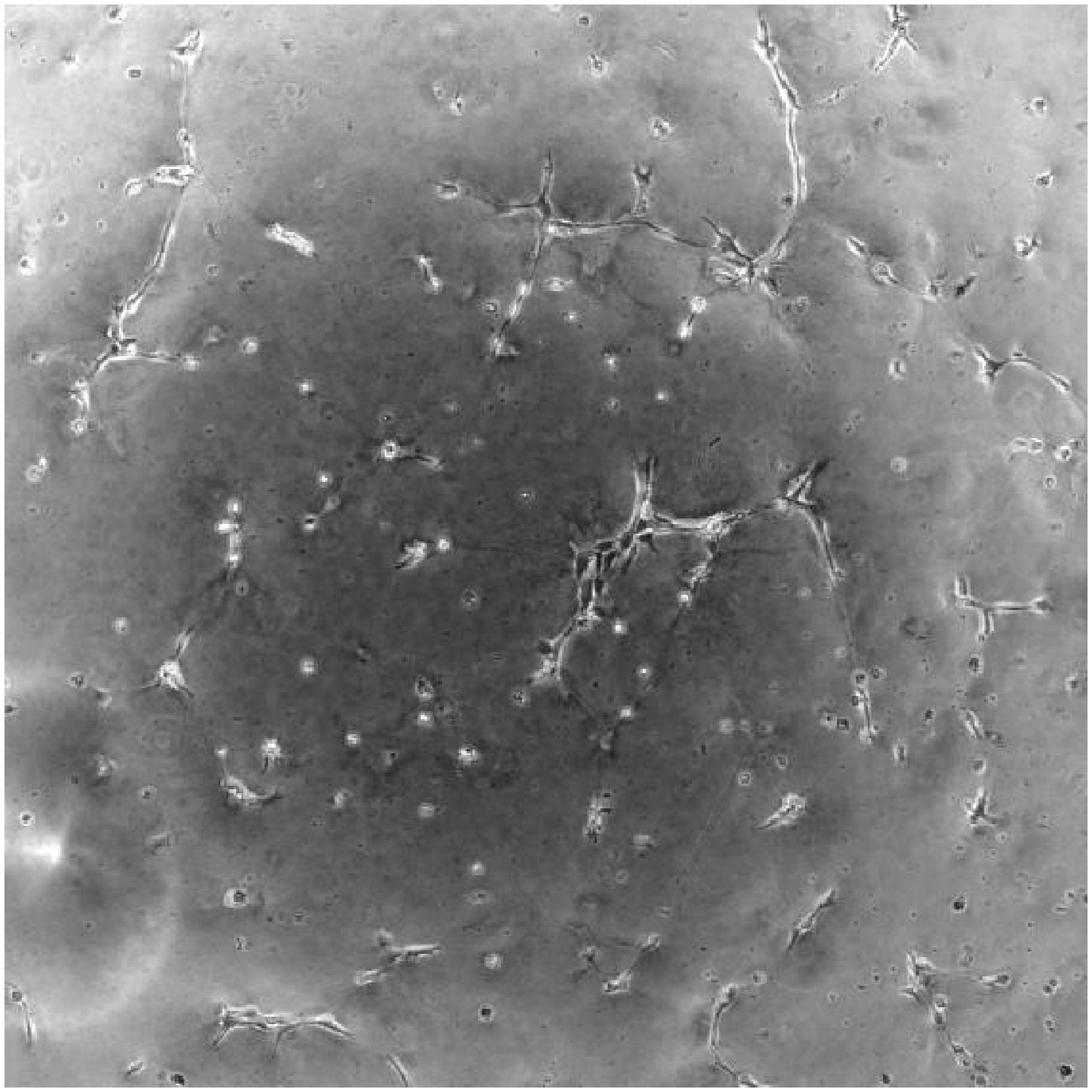}}&
b&\mbox{\epsfysize=2.5cm\epsfbox{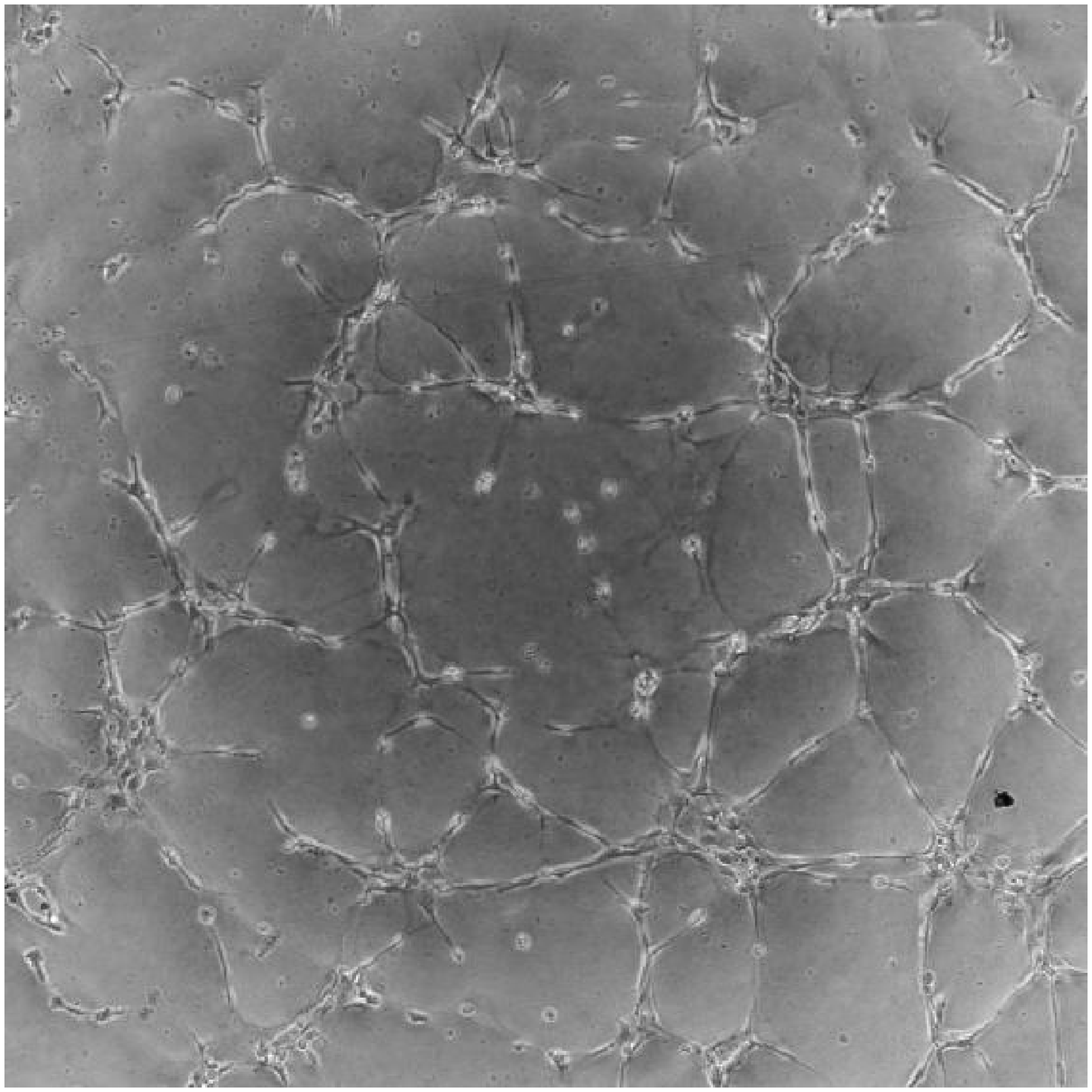}}\\
c&\mbox{\epsfysize=2.5cm\epsfbox{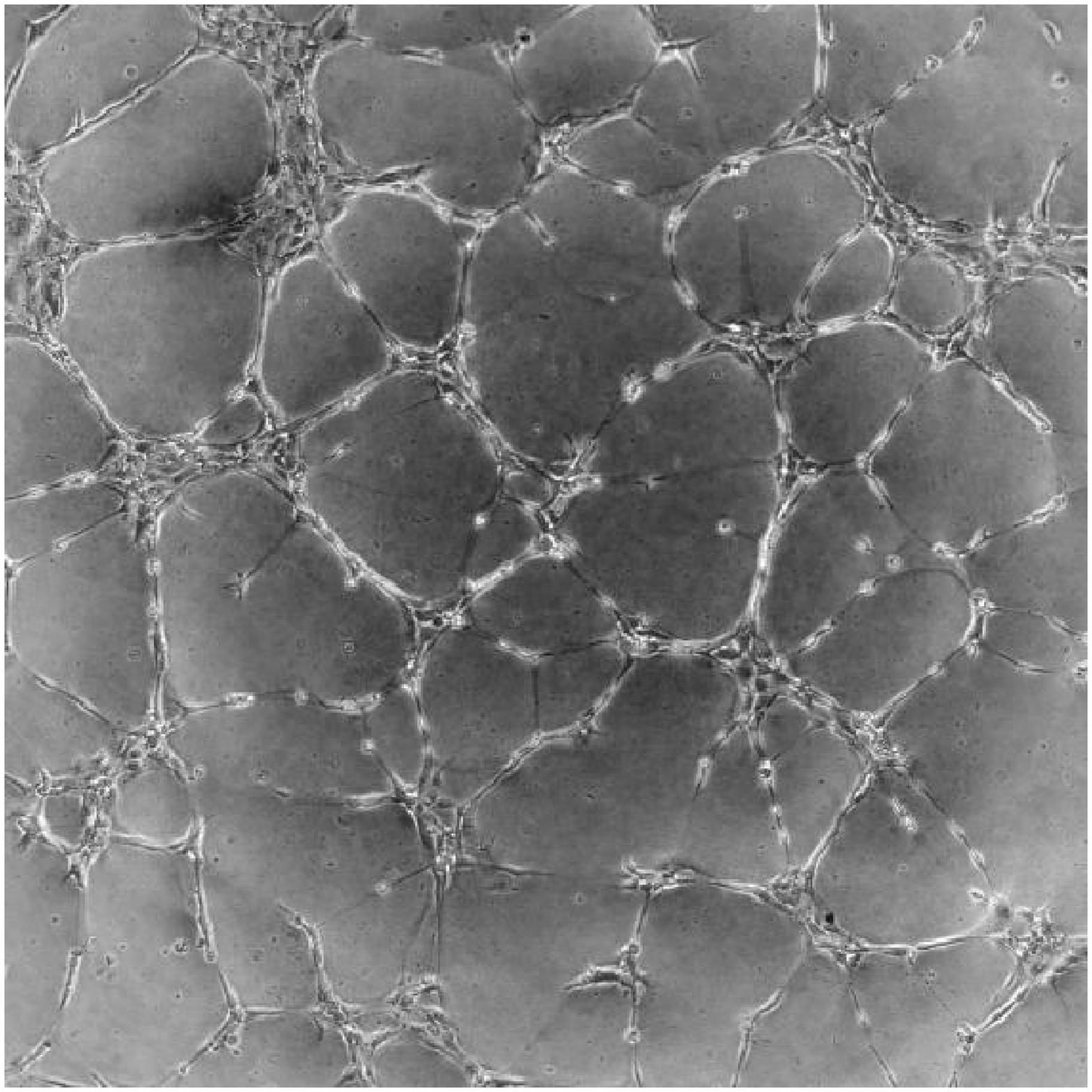}}&
d&\mbox{\epsfysize=2.5cm\epsfbox{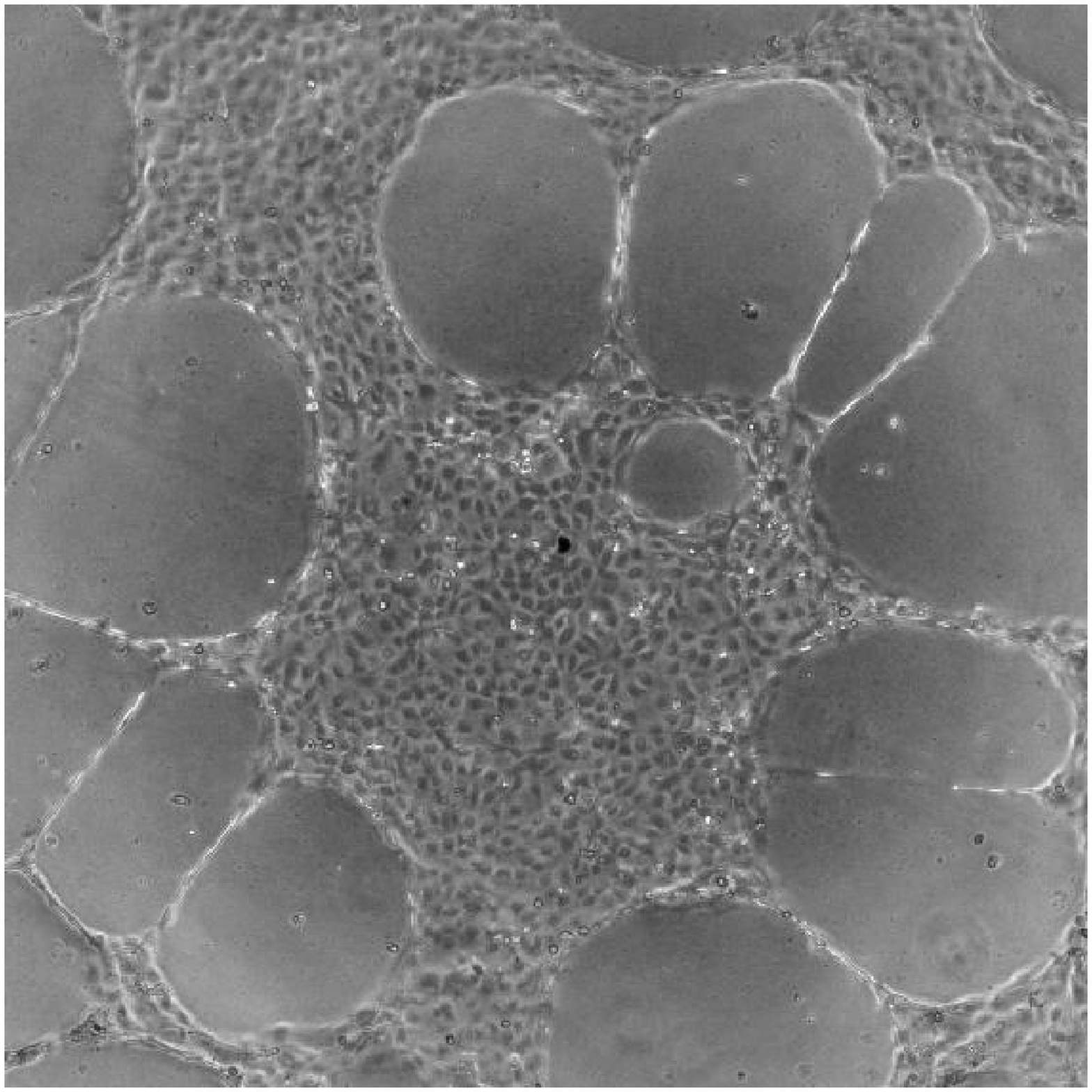}}\\
\end{tabular}
\ec
\caption{E\label{expict}xperimental pictures of vascular networks,
on a system of size $L=2$ mm,
obtained starting from four different
values of the initial cell density: a)  50 cells/mm$^2$;
b) 100 cells/mm$^2$;
c) 200 cells/mm$^2$;
d) 400 cells/mm$^2$.}
\end{figure}

The problem of morphogenesis and self-organization in biological systems is
both an actual and a long standing one \cite{yates88,jensen00}. In living
beings, complex structures arise from the aggregation of separate
constituents.
Such structures, although irregular, often show some sort of hidden order, like
self-similarity and scaling laws \cite{mandelbrot88,stanley2}.
Scaling laws are known
to emerge not only in the physics of phase transition \cite{stanley}
and in percolation \cite{sa94}, but also from several kinds of aggregation
dynamics, and are often considered as fingerprints of the process which led
to the formation of the structure itself \cite{vicsek92,vicsek01}.
A typical example of natural structure characterized by non trivial scaling
laws is the vascular network \cite{wbe}.
In recent years many experimental investigations have been performed on the
mechanism of blood vessel formation \cite{carmeliet00}. Cells are cultured
on a gel matrix and their migration and aggregation observed through
videomicroscopy.
Tracking of individual trajectories
shows marked persistence in the direction, with a small random component 
superimposed \cite{bio}.
The motion is directed towards zone of higher concentration
of cells, suggesting that chemotactic factors play a role.
Cells migrate over distances which are an order of magnitude larger than
their radius and aggregate when they get in touch with one of their neighbors.
In a time of the order of 10 hours they form a continuous multicellular
network which can be described as a collection of nodes connected by chords.
The mean chord length is approximately independent on the initial cell density
$\bn$, with an average value $\bar{\ell}= 200\pm 20\, \mu\mathrm{m}$ for $\bn$ between
100 to 200 cells/mm$^2$.
Observing the formed structures, one finds that,
by varying the initial cell density, there is a percolative transition.
Below a critical value $n_\mathrm{c}\sim 100$ cells/mm$^2$
groups of disconnected
structures are formed (Fig.\ \ref{expict}a). For $\bn\ge n_\mathrm{c}$ a single
connected network is instead visible (Fig.\ \ref{expict}b,c).
For higher values of $\bn$ one observes a sort of ``swiss cheese'' pattern
(Fig.\ \ref{expict}d).  In the biological system the percolating property
is of physiological relevance, since it is directly linked with the
functionality of blood vessels.

In this paper we propose a theoretical model which turns to be in
good agreement with experimental observations.
It allows to well reproduce both the observed percolative transition
and the typical scale of observed vascular networks.
With respect to standard percolative models, here an important role is
played by migration and dynamical aggregation of particles.
The model appears thus as a possibly new and physically
interesting representative in the class of percolation models. 
We characterize the phase transition from the point of
view of scaling laws, both in the model and in the real system, computing
critical indices and the fractal dimension of the percolating cluster
through extensive numerical simulations and analysis of the experimental data.
The model describes the cell
population as a continuous density field $n$ and velocity $\mathbf{v}$;
it also assumes the presence of a concentration field $c$ of soluble
factor.  Cells are modeled as a fluid
accelerated by gradients of the soluble factor. The latter is supposed to be
released by cells, diffuse, and degrade in finite time, in agreement with
experimental observations. These assumptions give rise to the following
equations:
\begin{subequations}
\label{burgers}
\begin{eqnarray}
&&\frac{\partial n}{\partial t}+\nabla \cdot \left( n\mathbf{v}\right)   =  0\label{continuita} \\
&&\frac{\partial \mathbf{v}}{\partial t}+\mathbf{v}\cdot \nabla \mathbf{v}  =  \mu \, \nabla \, c\label{riemann} \\
&&\frac{\partial c}{\partial t}  =  D\nabla^2 c+\alpha n-\tau ^{-1}c\label{diffusione} 
\end{eqnarray}
\end{subequations}
where \( D,\, \alpha ,\, \tau  \), are respectively the diffusion
coefficient, the rate of release and the characteristic degradation time of
soluble mediators, and \(\mu \) measures the strength of cell response.
Initial conditions are given in the form of a set of randomly distributed
bell-shaped bumps
in the density field, representing single cells initially spread in the
system, with zero velocities and zero concentration of the soluble factor.
In Eqs. (\ref{burgers}) a multidimensional 
Riemann (\ref{riemann}) equation is
coupled to the diffusion equation (\ref{diffusione}),
while (\ref{continuita}) is just a continuity equation expressing conservation
of the total cell number.
Multivaluedness of solutions to (\ref{riemann}) can be prevented by adding
a viscous term $\nabla^2\mathbf{v}$ in the r.h.s.
One than gets the analog of Burgers' equation, which has been
used to describe growth of molecular interfaces \cite{kpz86} and the
emergence of network-like patterns \cite{bik02} in the large-scale
distribution of masses of the Universe \cite{z70,vdfn94}.
Instead of viscosity, we introduce a
phenomenological, density dependent pressure term $-\nabla \phi(n)$,
in the r.h.s. of (\ref{riemann}), 
with $\phi(n)$ zero for low densities, 
and rapidly increasing for densities above a threshold $\sigma^{-2}$,
where $\sigma=30\,\mu\mathrm{m}$ is the mean cell radius,
in order to describe the fact that cells do not
interpenetrate.

The model provides a natural lenght scale $r_{0}=\sqrt{D\tau}$,
which is the effective range of the
interaction mediated by the soluble factor.
This suggests that, starting from the above
mentioned initial conditions, Eqs.\ (\ref{burgers}) should develop
network patterns characterized by the same scale.
Direct measurements of $D$ and $\tau$ give $D\sim 10^{-7}$ cm$^2$/s,
$\tau \sim$ 1h, from which one gets $r_{0}\simeq$ 200 $\mu$m, a value
which is in good agreement with the mean chord length $\bar\ell$
experimentally observed.

\begin{figure}
\bc
\begin{tabular}{cccc}
a&\mbox{\epsfysize=2.5cm\epsfbox{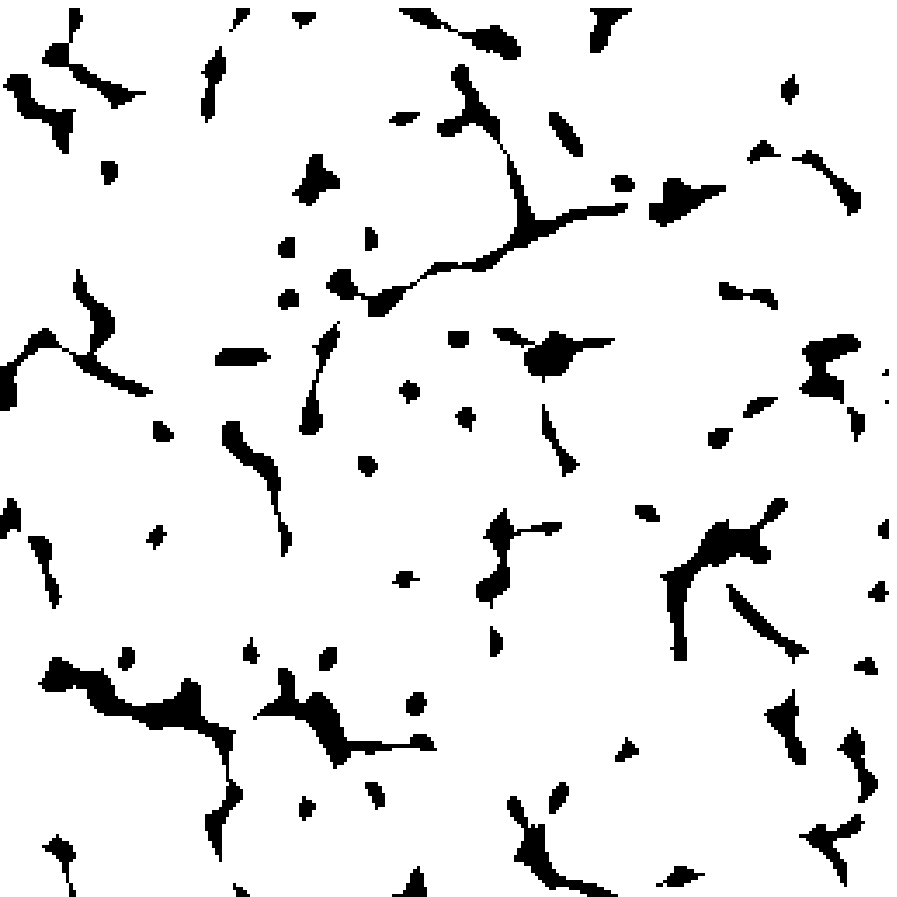}}&
b&\mbox{\epsfysize=2.5cm\epsfbox{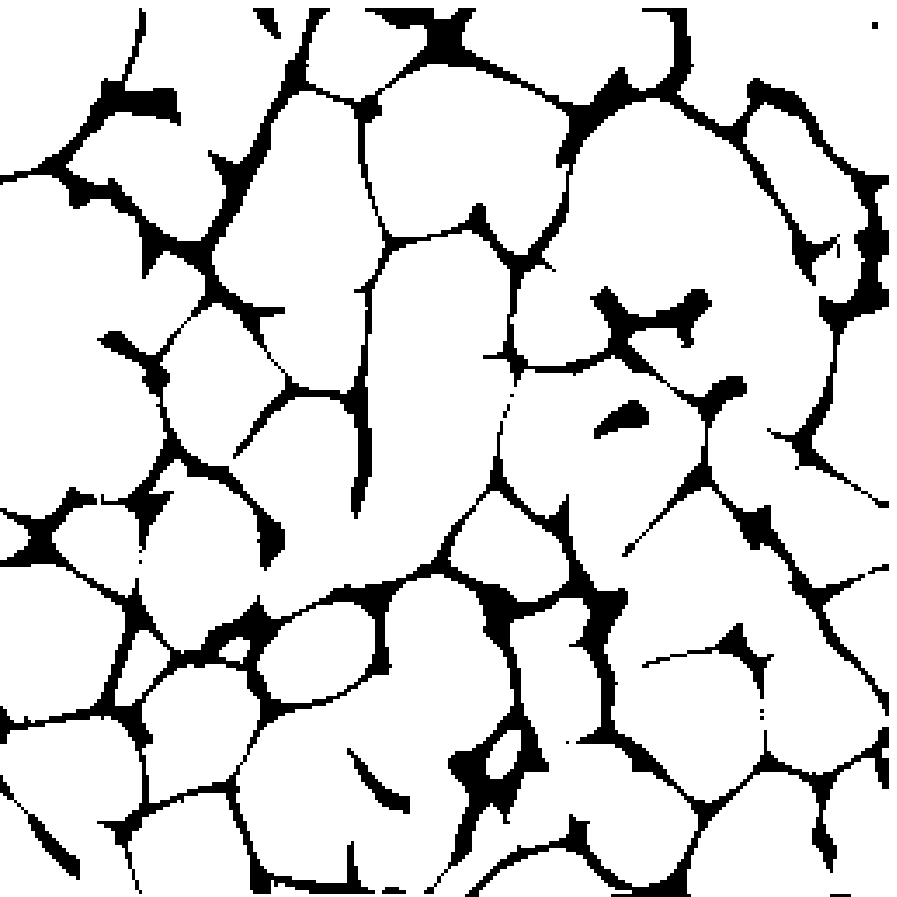}}\\
c&\mbox{\epsfysize=2.5cm\epsfbox{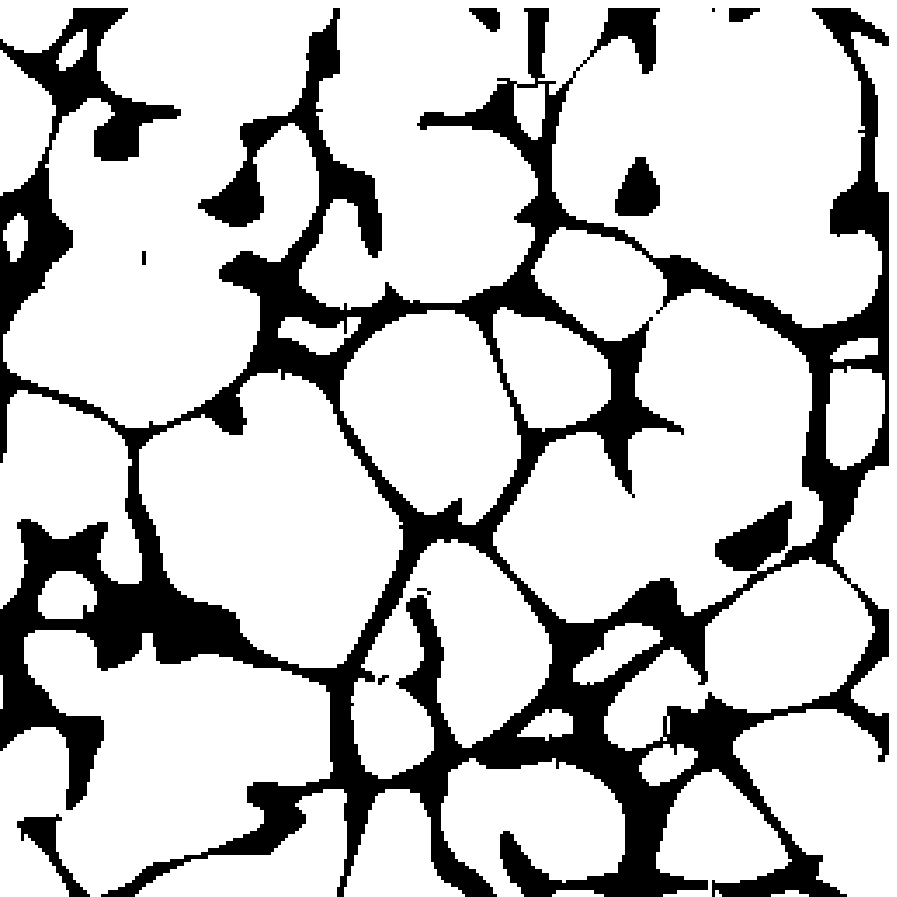}}&
d&\mbox{\epsfysize=2.5cm\epsfbox{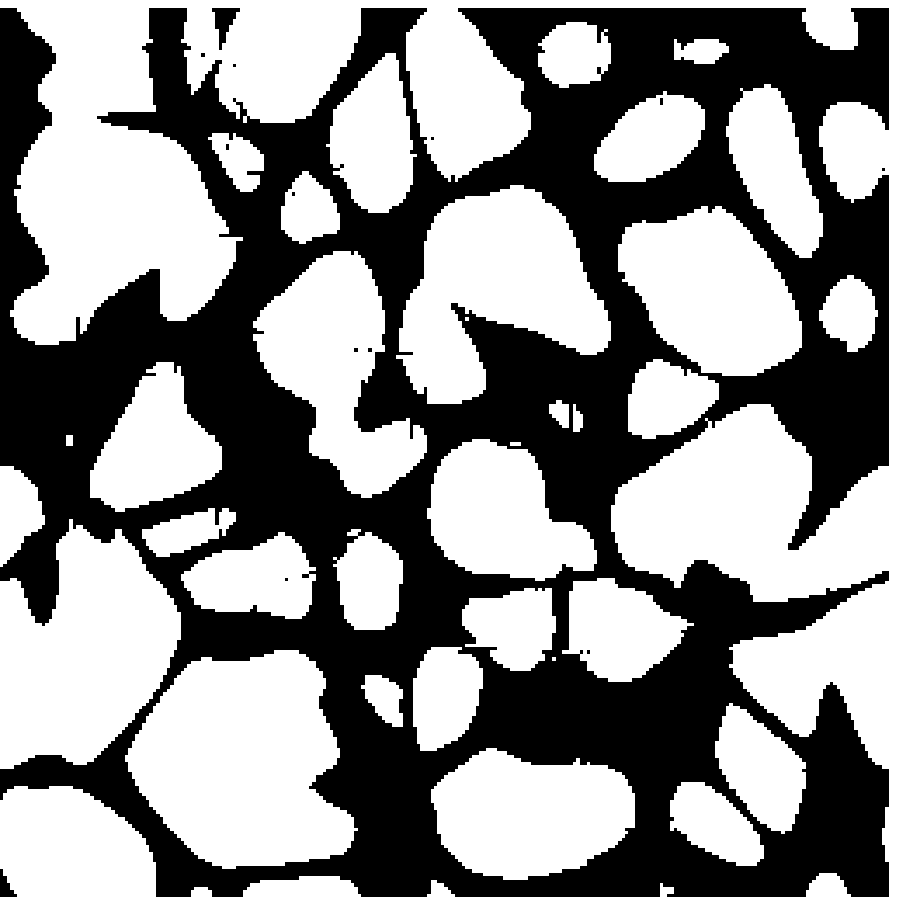}}\\
\end{tabular}
\ec
\caption{\label{simulpic}Numerical simulations, on a system of size $L=2$ mm,
obtained starting from four different values of the initial
cell density: a)  50 cells/mm$^2$;
b) 100 cells/mm$^2$;
c) 200 cells/mm$^2$;
d) 400 cells/mm$^2$.}
\end{figure}

We performed numerical simulations of model~(\ref{burgers})
on square boxes of size $L=1$, 2, 4, $8\,{\rm mm}$, with
periodic boundary conditions,
using a finite volume method~\cite{leveque}.
We used the 
experimental
values for \(D\) and \(\tau\),
while the unknown parameters \(\mu,\,\alpha\) were set to one
(which amounts to an
appropriate rescaling of time and the concentration field \(c\)).
The initial conditions were given by throwing the same number 
of cells as in the biological experiments 
in random positions inside the box, with zero velocities
and zero concentration of the soluble factor, with a single
cell given initially by a Gaussian bump of width $\sigma$
and unitary weight in the integrated cell density field \(n\).
Starting from these condition, Eqs. (\ref{burgers}) 
were numerically integrated. The simulation was stopped when the
vascular network was formed, or a stationary state was reached.
We observed a remarkable
correspondence between simulations and experiments both in the
the dynamic evolution details (not shown) and in the reproduction of the
percolating patterns (Fig.~\ref{simulpic}). 

In order to study the percolative transition
we computed between 100 to 200 realizations of the system for
each box size and cell density, with different initial positions of
the cells.
The simulation gives as a result a continuous density
field $n$. In order to study percolation properties we partitioned the box
in little square boxes of size $2^{-6}$ mm,
forming the sites of a square
lattice, and set each site as filled if the mean cell density inside the
corresponding little box was larger than the threshold $\sigma^{-2}$, empty
otherwise.
We then identified clusters of nearest neighbor filled sites.
In Fig. \ref{simulpic} the results of four simulations, for different values
of the density, are shown. Black areas represent regions filled with cells,
that is regions where density exceeds the threshold, white areas represent
the underlying substrate.
As in the case of experimental data, a percolative transition is
observed at some critical cell density $n_\mathrm{c}$.
Furthermore the model reproduces
quite well the typical structure of the vascular network, with chords of
length $\bar{\ell}\simeq$ 200 $\mu$m for a wide range of cell densities,
in good agreement with the expected value $r_0$.
\begin{figure}
\bc
a\mbox{\epsfysize=3.2cm\epsfbox{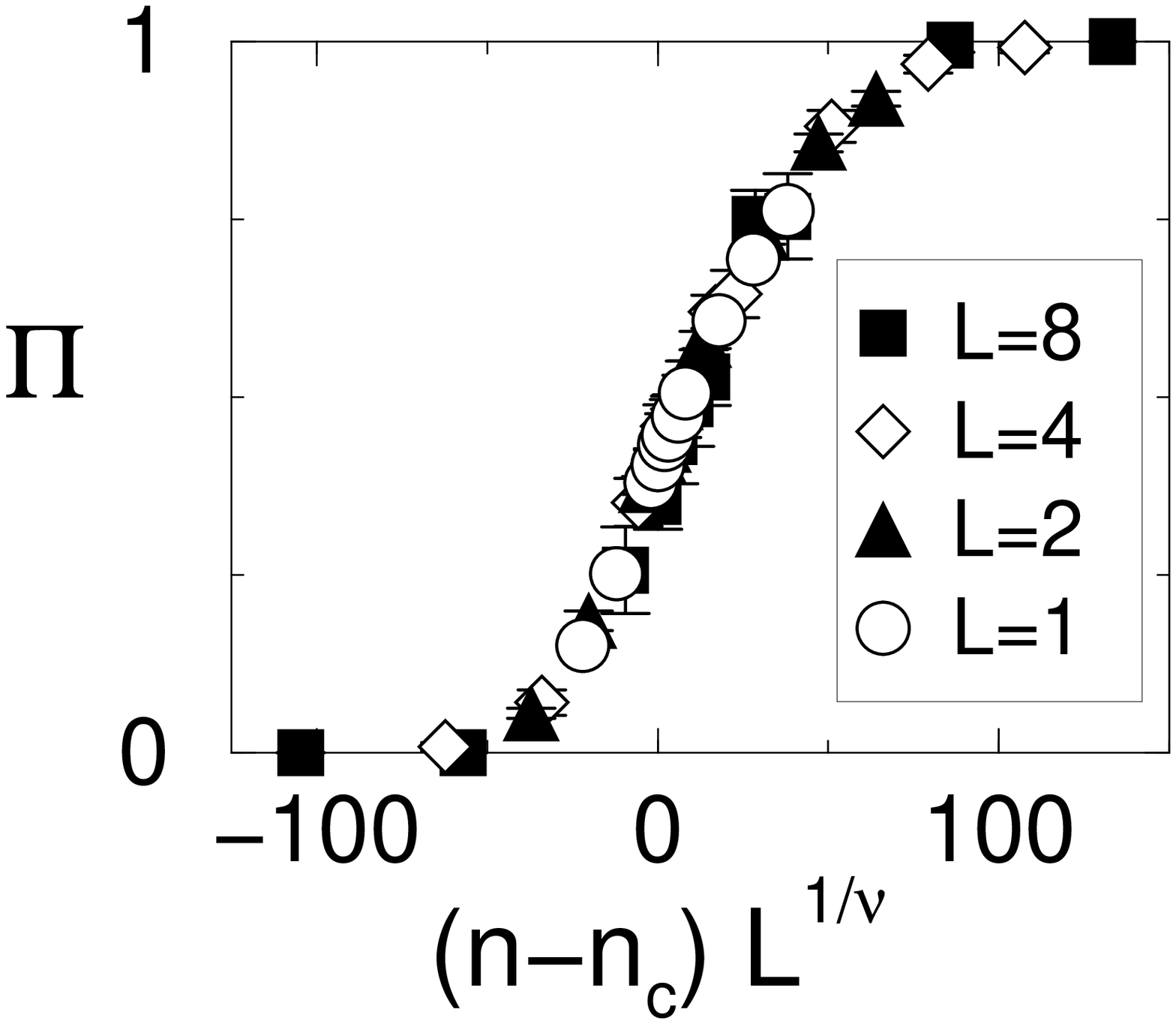}}
\quad b\mbox{\epsfysize=3.2cm\epsfbox{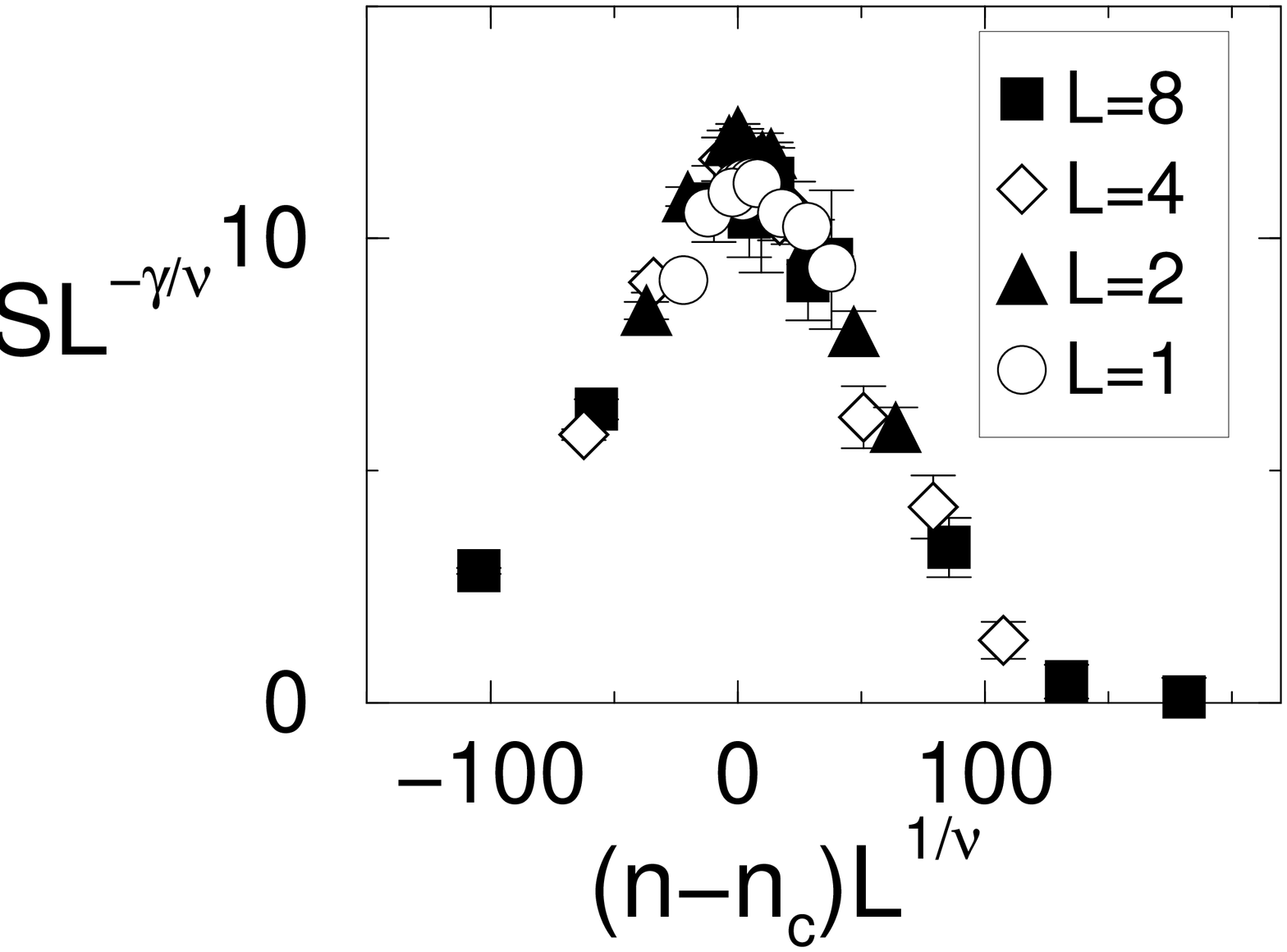}}\\
\ec
\caption{\label{collapse}Data collapse of a) percolation probability $\Pi$ and
b) mean cluster size S,
for the numerical simulations with system sizes $L=1$, 2, 4, and 8 mm.}
\end{figure}

For each set of realizations, corresponding to a value of the box size $L$
and of the mean cell density $\bn$, we identified clusters of nearest
neighbor filled sites, and measured the following quantities:
1) the percolation probability $\Pi$, defined as the fraction of realizations
in which a percolating cluster appears; 2) the infinite cluster density $P$,
defined as the density of sites belonging to the infinite (greatest) cluster;
3) the mean
cluster size $S=(1/N)\sum_s n_s s^2$, where $N$ is the total number of sites
and $n_s$ is the number of clusters of size $s$, excluding from the sum the
infinite cluster.

In percolation models, in presence of a second order transition at some
mean cell density $n_\mathrm{c}$, these quantities show a characteristic singular
behavior near the percolation point.
Namely, the percolation probability is zero below the transition and
one above it; the infinite cluster density, which plays the role of
order parameter, vanishes as $|\bn-n_\mathrm{c}|^\beta$ from above; and
the mean cluster size has a singularity $|\bn-n_\mathrm{c}|^{-\gamma}$ on both
sides of the transition.
For finite size systems, the same quantities obey near the transition to
the following relations:
\begin{subequations}
\label{fss}
\begin{eqnarray}
\label{pifss}
\Pi(\bn,L) &\>\sim\>&  \widehat{\Pi}[(\bn-n_\mathrm{c})L^{1/\nu}]\\
\label{pfss}
P(\bn,L)   &\sim& L^{-\beta/\nu}\, \widehat{P}[(\bn-n_\mathrm{c})L^{1/\nu}]\\
\label{sfss}
S(\bn,L)   &\sim& L^{\gamma/\nu}\, \widehat{S}[(\bn-n_\mathrm{c})L^{1/\nu}]
\end{eqnarray}
\end{subequations}
Using Eq. (\ref{pifss}), one can find the critical density $n_\mathrm{c}$ as the
point where the curves $\Pi(\bn,L)$ for different sizes $L$ cross.
The estimated value is $n_\mathrm{c}=94\pm 1$ cells/mm$^2$.
Then, the critical index $\nu$ is found by plotting $\Pi(\bn,L)$ as a function
of $(\bn-n_\mathrm{c})L^{1/\nu}$, where $n_\mathrm{c}$ is held fixed to the value previously
computed, and looking for the value of $\nu$ that gives the
best data collapse. The error on the computed value of the index is estimated
looking for the range of values that gives an acceptable data collapse.
Holding the value of $n_\mathrm{c}$ and of $\nu$ fixed, the other two relations
Eq.\ (\ref{pfss}) and (\ref{sfss}) are used to estimate the value of the 
indices $\beta$ and $\gamma$, by looking for the data collapse of
$L^{\beta/\nu}\,P(\bn,L)$ and $L^{-\gamma/\nu}\,S(\bn,L)$ as a function of
$(\bn-n_\mathrm{c})L^{1/\nu}$. 
We observed that while \(n_\mathrm{c}\) is sensitive to the choice of 
the threshold in a neighborhood of the natural value \(\sigma^{-2}\),
the critical exponents do not depend on it.
In Fig.\ \ref{collapse} the data collapses
for $\Pi$ and $S$ are shown. The estimated values of the critical indices
are reported in the second column of Tab.\ \ref{tabindices}.
\begin{table}
\bc
\begin{tabular}{|c|c|c|c|}
\hline
critical index&
model&
experiments&
random percolation\\
\hline
$\nu$        & $1.33\pm 0.08$ & not measured   & $1.333$\\     %
$\gamma/\nu$ & $1.83\pm 0.05$ & $ 1.78\pm 0.12$ & $1.792$\\   %
$\beta/\nu$  & $0.11\pm 0.01$ & not measured & $0.104$\\     %
$D$          & $1.87\pm 0.03$ & $1.85\pm 0.10$  & $1.896$\\   %
\hline
\end{tabular}
\ec
\caption{\label{tabindices}Critical indices measured on the numerical model and on
experimental data, and compared with the exact values of random percolation.}

\end{table}

\begin{figure}
\bc
\mbox{\epsfysize=5.0cm\epsfbox{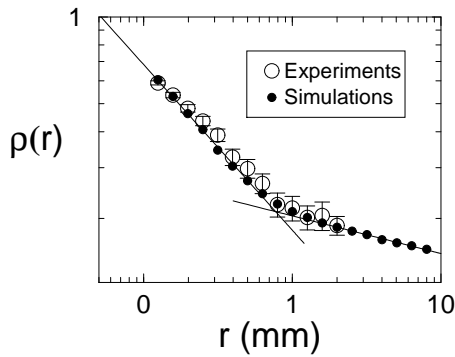}}
\ec
\caption{\label{fractdim}Density of the percolating cluster as a function of the radius
for numerical simulations and experimental data.}
\end{figure}

Within the errors, the values obtained are the same that one expects for the
simple random percolation model, that is when the sites of the lattice are
uncorrelated, and each one occupied with the same probability.
As critical indices are linked to the large scale structure of the percolating
cluster, this means that, on such a large scale, the structure of the vascular
network is mainly determined by the initial random positioning of the cells,
and is not altered by the dynamical process of migration and aggregation.
On the other hand, on smaller distances clusters are quite different from
the ones of random percolation. For example the mean fraction of occupied
sites, which is a quantity sensible to the small scale structure of clusters,
is about 0.2 at the percolation point, much lower than the corresponding
figure for random percolation, that is 0.59. 
A quantity that can give us informations about the structure of the percolating
cluster at different scales is the density $\rho(r)$
as a function of the radius.
This is defined as the mean density of sites belonging to the percolating
cluster, inclosed in a circle of radius $r$ centered at one site belonging
to the cluster, and averaged over different centering sites and different
realizations of the system. For a fractal object
with fractal dimension $D$,
this should scale as $\rho(r)\sim r^{D-d}$.
For the percolating cluster of random percolation at the critical point,
one expects a fractal dimension $D=1.896$.
We computed $\rho(r)$ for our model at the percolation point
$\bn=n_\mathrm{c}$ (see Fig.\ \ref{fractdim}),
and found two different regimes depending on the scale. 
For $r>r_\mathrm{c}$, with $r_\mathrm{c}=0.77\pm 0.08$ mm, a
behavior compatible with random percolation is found, with fractal
dimension $D=1.87\pm 0.03$. On the other hand, for $r<r_\mathrm{c}$ a different
exponent is found, $D=1.50\pm 0.02$. 
This exponent may be the signature of the dynamic growth process driven for
$r<r_\mathrm{c}$ by the rapidly oscillating components of the concentration field.

We have repeated the measures described above on 
a set of experimental data,
consisting of 28 digital photographs (with a resolution of
1024$^2$ pixels) of mature structures obtained on a matrix of size $L=2$ mm,
starting from initial densities of 50, 75, 100, $\ldots$, 200 cells/mm$^2$.
The photographs are the results of two experimental sessions, each performed
in duplicate, so that 4 photographs are available for each density value.
The data show the presence of a percolative transition at $n_\mathrm{c}\sim 125$
cells/mm$^2$, and are compatible with the values %
$\gamma/\nu=1.83$
obtained from numerical simulations.   The available data are
not enough to obtain $\nu$ and $\beta/\nu$ with a reasonable precision.  
Estimated values of the critical
indices are reported in the third column of Tab.\ \ref{tabindices}.
Finally, we measured the density of the percolating cluster as a function
of the radius at the percolation point,
and found a behavior surprisingly similar to the one found in
the numerical model (see Fig.\ \ref{fractdim}):
for $r>r_\mathrm{c}$ the fractal dimension of the percolating
cluster is $D=1.85\pm 0.10$ which is compatible with $D=1.87\pm 0.03$
from the numerical simulations and with the value $D=1.896$ of random
percolation. For $r<r_\mathrm{c}$ instead $D=1.48\pm 0.05$.

In conclusion, we have introduced a model which describes the formation
of the vascular network with a typical scale lenght $r_0$,
and is in good agreement with
experimental data.
The percolative phase transition appears to be second
order, at least to the accuracy allowed by the limitations in size which are
intrinsic to our numerical experiments. Comparison of critical indices
suggests that the phase transition falls in the universality class of random
percolation, even in presence of migration and dynamical aggregation.
This is confirmed by the fractal dimension of the percolating cluster,
on scales larger than $r_\mathrm{c}\simeq 0.8$ mm, that is in agreement
with the one expected for random percolation.
On the other hand we found that, if observed on scales smaller than $r_\mathrm{c}$,
the vascular network shows a different scaling behavior,
with a fractal dimension $D\simeq 1.5$, that may be the signature of the
dynamical process that led to the formation of the network.

Thus, the model appears to be quite successful in describing
{\em in vitro} experiments, where all the parameters are under control
and one can easily tune the cell density. 
It would also be interesting to understand the relevance of this picture
to vascular networks in living beings.
It is known \cite{carmeliet00} that during embriogenesis blood
vessels form through local ({\em in situ}) differentiation of a fraction
of endothelial cells, that afterwards assemble in a vascular labyrinth,
a process similar to the one described in this paper. 
However, in this case it is more difficult to have complete control over
all the relevant parameters. In particular, 
one would expect that the cell density is fixed through some feedback
mechanism, in order to maximize efficiency \cite{wbe}.
Indeed, some experiments \cite{gblb95} on {\em in vivo} capillaries 
show compact space-filling structures, with fractal dimension $D=2$,
which would indicate that cell density is tuned by Nature to a value
larger than the percolation threshold.
More experiments, %
focusing on the relevance of the percolation
mechanism, and the nature of cell density self-regulation,
are needed to shed light on these issues.

Acknowledgments --
The authors acknowledge useful discussions 
with N.~Bellomo, I.~Kolokolov, P.~Netti
and M.~Vergassola.
This work was partially supported by the European TMR Network-Fractals
(Contract No. FMRXCT980183), RTN HPRN-CT-2000005,
CNR 00.00141.ST74, MURST-PRIN-2000, 
MURST-Cofin-2001, INFM-PRA (HOP), and MIUR-FIRB-2002.

\end{document}